# IP Geolocation Underestimates Regressive Economic Patterns in MOOC Usage

Daniela Ganelin
Massachusetts Institute of Technology
77 Massachusetts Ave.
Cambridge, MA 02139
dganelin@alum.mit.edu

Isaac Chuang
Massachusetts Institute of Technology
77 Massachusetts Ave.
Cambridge, MA 02139
ichuang@mit.edu

## ABSTRACT

Massive open online courses (MOOCs) promise to make rigorous higher education accessible to everyone, but prior research has shown that registrants tend to come from backgrounds of higher socioeconomic status. We study geographically granular economic patterns in ~76,000 U.S. registrations for ~600 HarvardX and MITx courses between 2012 and 2018, identifying registrants' locations using both IP geolocation and user-reported mailing addresses. By either metric, we find higher registration rates among postal codes with greater prosperity or population density. However, we also find evidence of bias in IP geolocation: it makes greater errors, both geographically and economically, for users from more economically distressed areas; it disproportionately places users in prosperous areas; and it underestimates the regressive pattern in MOOC registration. Researchers should use IP geolocation in MOOC studies with care, and consider the possibility of similar economic biases affecting its other academic, commercial, and legal uses.

## 1. INTRODUCTION

Since 2012, millions of users worldwide have enrolled in massive open online courses (MOOCs) offered by Harvard University and MIT on edX [1]. MOOC developers initially hoped to "democratize and reimagine education so that anyone, anywhere, regardless of his or her social status or income, can access education" [2]. However, studies have found that MOOC users tend to come from backgrounds of higher socioeconomic status, raising doubts that MOOCs are providing access to education to those who could not otherwise afford it. Both registrants [3] and course completers [4] disproportionately come from developed countries. MOOC users report having high levels of education [2, 3], particularly compared to peers in developing countries [5]. Studies find that within both India [6] and the U.S. [7], registrants come from areas of high prosperity and population density.

One difficulty facing researchers is that they generally have no direct measures of users' economic backgrounds. As a workaround, some researchers attempt to identify users' locations, so that they can use a geographic area's socioeconomic status as a proxy for the individual's. Location identification can rely on users' self-reported mailing addresses, as in [7], or their devices' recorded Internet Protocol (IP) addresses, as in [6].

IP geolocation - the process of identifying users' locations from IP addresses – is tempting for researchers as an easily scalable, relatively inexpensive tool. It relies on commercial databases which tie sets of IP addresses to granular physical locations.

Unfortunately, IP geolocation can fail in many ways. Internet service providers can arbitrarily reassign IP addresses; a registrant using dialup Internet, a virtual proxy network, Secure Shell, or Tor can present with an IP address from a device located far away from her; databases, which do not publicize their sources, can contain outdated or simply erroneous entries [8]. Indeed, studies from the networking community find that IP geolocation is substantially inaccurate: "there is a long and fat tail of errors in the databases… in the range of thousands of kilometers and countries apart" [9]. "In most of the cases however, the location given by the databases is off by several hundreds, even thousands of kilometers"; "Geolocation databases can claim country-level accuracy, but certainly not city-level" [10].

However, MOOC studies using IP geolocation seldom mention concerns about the accuracy of the technology. Furthermore, when studying economic questions, there is the possibility that IP geolocation has not just limited accuracy, but also systematic bias – for example, that it is more accurate in places with greater prosperity and population density, where Internet infrastructure is more developed [11].

In this paper, we study economic patterns in U.S. HarvardX and MITx registrations using registrant locations, according to IP geolocation and to self-reported mailing addresses. Consistently with previous research, we find a regressive economic pattern by either metric: registrants disproportionately come from wealthier, denser ZIP Codes. We also find that compared to mailing addresses, which we treat as ground-truth data, IP geolocation is economically biased. It underestimates the regressive pattern, errs more for users from economically distressed areas, and disproportionately places users in more prosperous areas. We conclude that MOOC and other researchers should use IP geolocation with caution and note the possibility of bias. A longer version of this work is available at [12].

## 2. METHODOLOGY
### 2.1 Summary of Datasets

This work relies on combining the following sources of data:

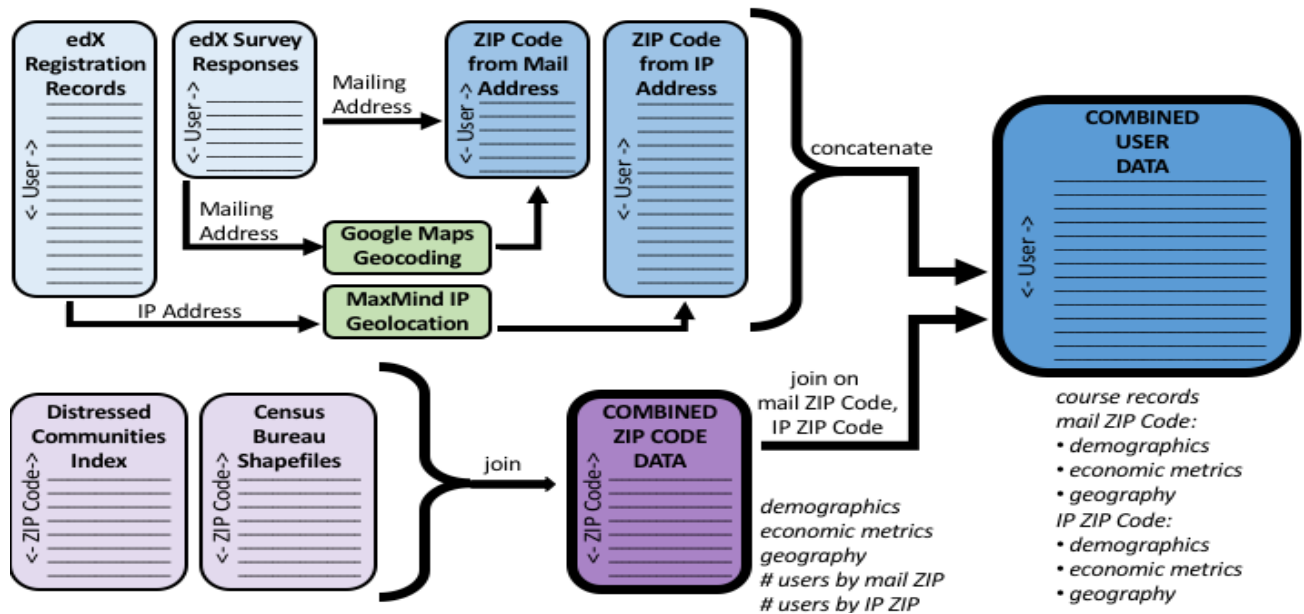
**Figure 1. Combining datasets**

- *IP addresses from EdX.* We use records for U.S.-identified users who registered for HarvardX and MITx courses on EdX between 2012 and early 2018, regardless of future course participation or completion. The records include a modal IPv4 address for each user, for each enrolled course. We use the IP address from each user's earliest course enrollment.
- *Mailing addresses from EdX.* ~135,000 users answered a survey question asking for a mailing address, although many responses are incomplete, outside the U.S., or not addresses.
- *MaxMind IP Geolocation database [13].* The MaxMind GeoIP2 City database maps sets of IP addresses to postal codes. It claims to identify the correct U.S. ZIP code with 36% accuracy.
- *Google Maps Geocoding API [14].* The Google Maps Geocoding API translates an address, which may be imperfectly formatted, to a well-formatted address, ZIP Code, and geographic coordinates.
- *Distressed Communities Index (DCI) [15].* The Distressed Communities Index dataset provides demographic and economic data for most ZIP Codes. It includes DCI, a summary score of economic distress in percentiles from 0 (most prosperous) to 100 (most distressed).
- *ZIP Code Shapefiles [16].* The U.S. Census Bureau distributes files which give an area and geometry (a list of coordinates forming a polygon) for each Zip Code Tabulation Area[1].

Figure 1 shows how we combine the datasets. Starting from edX user data, we use parsing and Google Maps Geocoding to obtain a ZIP Code from each mailing address, and MaxMind to obtain a ZIP Code from each IP address. For each user's two identified ZIP Codes, we join in demographic and economic data from DCI and geographic data from shapefiles.

## 2.2 Identifying Locations and Calculating IP Geolocation Error

For each user-provided mailing address, we attempt to extract a "ground-truth" ZIP Code. We remove invalid responses and parse addresses. If an address includes either a ZIP Code or a city and state, we use Google Maps Geocoding to obtain geographic coordinates and, if needed, a missing ZIP Code. After filtering based on the API's reported precision, we identify a "ground-truth" U.S. ZIP Code for ~79,000 users.

We use MaxMind to geolocate these users' IP addresses to "GeoIP" ZIP Codes. We obtain both a ground-truth ZIP Code and a GeoIP ZIP Code for ~76,000 users, whom we focus on in all analyses. We use latitude/longitude coordinates for ~73,000 of these. We join the user data with the Census Bureau shapefiles and the DCI economic and demographic dataset.

For the ~69,000 users where data are available, we calculate a metric of IP geolocation error that we call "boundary distance": the Euclidean distance between the user's ground-truth coordinates and the nearest point in the GeoIP ZIP Code. If the two ZIP Codes are identical, this distance is 0.[2,3]

## 2.3 Analyses

~16,000 ZIP Codes appear in ground-truth or GeoIP identifications. Where data are available, we tier them into deciles by population density, area, and population; in some analyses, we combine adjacent deciles into quintiles. We also consider ZIP Codes according to DCI tier (0 to 20 is Prosperous, …, 80 to 100 is Distressed), and analogously define ten sub-tiers (0 to 10, 10 to 20, …, 90 to 100).

---

[1] ZIP Code Tabulation Areas, geographic areas publicized by the U.S. Census Bureau, are technically distinct from ZIP Codes, the U.S. Postal Service's proprietary sets of mailing addresses. The DCI dataset uses them interchangeably, and so do we.

[2] Although this error is calculated in degrees latitude/longitude, for interpretability we show results in approximate miles; in the U.S., one degree is very roughly 50 miles.

[3] As a sensitivity check, we repeat analyses using an alternative geolocation error: great-circle distance between the internal points of the two ZIP Codes. We find very similar results.

We perform graphical and descriptive analyses to study economic patterns in registration and in geolocation error as well as patterns of GeoIP ZIP code properties relative to ground-truth ZIP Code properties, focusing on population density and DCI. We also compare overall DCI distributions according to ground-truth and GeoIP ZIP Codes.

## 3. RESULTS
### 3.1 Economic Patterns in Registrations

According to both ground-truth and GeoIP identifications, per-capita MOOC registration rates generally increase with ZIP Code population density, and decrease with area and level of distress (Figure 2). In absolute terms, too, the bulk of users come from more populous, lower-area, denser, lower-DCI ZIP Codes.

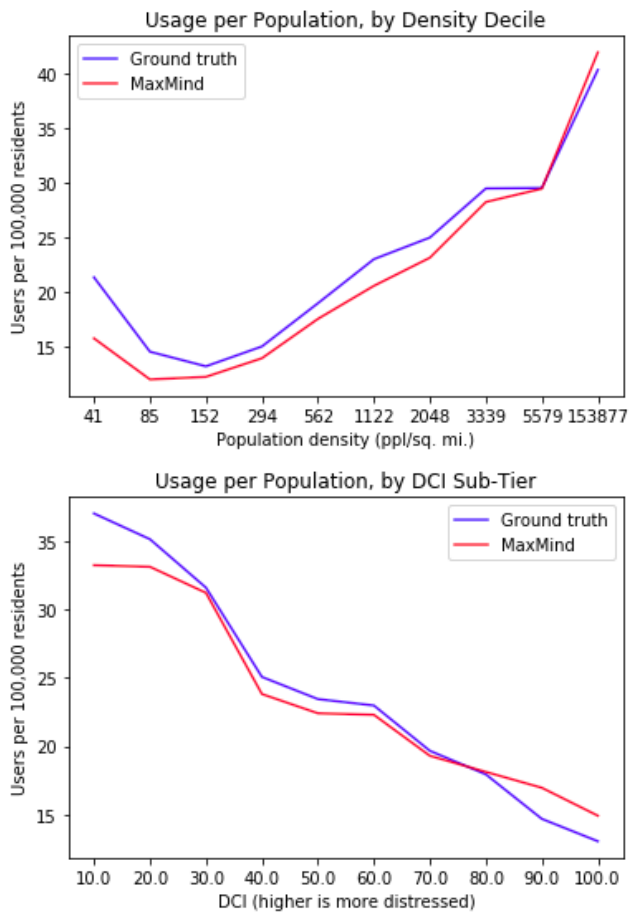

Figure 2. (a) Per-capita usage increases with ZIP Code density (b) Per-capita usage decreases with ZIP Code DCI

### 3.2 Economic Patterns in Geolocation Error

In assessing geolocation accuracy, there are two broad considerations: the probability that IP geolocation identifies the correct ZIP Code, and the magnitude of the error when it does not.

Overall, the GeoIP ZIP Code matches the ground-truth one for 18% of users, half of MaxMind's advertised number. The probability of an exact match increases with ground-truth population and ground-truth area (Table 1, Column 3). The latter makes sense intuitively: landing in the correct ZIP Code seems easier if that area is physically large.

When errors do occur, on average their magnitude decreases with population and density, and increases with DCI and area (Table 1, Column 4). These differences are most noticeable at the tail of the error distributions – e.g. errors at the 20$^{th}$ percentile are about the same across density tiers, but 90$^{th}$ percentile errors are much greater for lower-density ZIP Codes. The colored curves in Figure 3 show percentiles of geolocation error, from 10th at left to 90th at right, for each ZIP Code tier.

**Table 1. IP geolocation accuracy by ground-truth ZIP Code properties**

(1) Defining property and unit of ZIP Code tier.
(2) Upper cutoff of ZIP Code tier, according to units of (1).
(3) Probability of exact-ZIP match: among users with ground-truth ZIP Code in this tier, percentage where GeoIP ZIP Code matches ground-truth ZIP Code.
(4) Average geolocation error (approximate miles): among users with ground-truth ZIP Code in this tier *without* an exact-ZIP match, exponentiated mean of logarithm of boundary distance.
(5) Spread of geolocation error (approximate miles): Among same users as (4), exponentiated standard deviation of logarithm of boundary distance.

| (1) | (2) | (3) (%) | (4) (mi.) | (5) (mi.) |
|---|---|---|---|---|
| Population (thousands) | 115 | 23.0 | 8.3 | 10.2 |
| | 33 | 17.7 | 7.5 | 11.0 |
| | 21 | 13.7 | 7.6 | 11.0 |
| | 12 | 10.6 | 9.8 | 10.0 |
| | 5 | 7.6 | 13.5 | 10.1 |
| Area (sq. mi.) | 7,750 | 27.3 | 25.4 | 9.1 |
| | 94 | 22.0 | 15.1 | 9.4 |
| | 39 | 20.3 | 10.6 | 9.1 |
| | 16 | 17.3 | 7.6 | 9.7 |
| | 6 | 15.0 | 4.7 | 11.5 |
| Population density (people/sq. mi.) | 153,877 | 16.9 | 5.5 | 11.1 |
| | 3,339 | 19.0 | 8.4 | 9.6 |
| | 1,122 | 21.5 | 12.8 | 9.5 |
| | 294 | 21.1 | 18.3 | 8.8 |
| | 85 | 17.2 | 29.8 | 7.8 |
| DCI (percentile) | 100 | 18.2 | 9.1 | 11.4 |
| | 80 | 18.9 | 8.3 | 11.5 |
| | 60 | 19.3 | 8.6 | 11.1 |
| | 40 | 18.2 | 8.1 | 11.1 |
| | 20 | 18.6 | 8.0 | 9.6 |

### 3.3 Bias in ZIP Code Property Identification

A user's chances of being geolocated to a ZIP Code of roughly correct population, area, density, and DCI depends on his ground-truth ZIP Code. Figure 4 shows the distribution of GeoIP ZIP Code density and DCI for users from particular ground-truth ZIP Code tiers. In each case, GeoIP ZIP Codes are disproportionately drawn to one end of the distribution, and users from that favored tier are more likely to be placed to the correct tier.

For example, geolocation is more successful for ZIP Codes with lower DCI. A user from a Prosperous ZIP Code has a 59% chance of being geolocated to a Prosperous ZIP Code, while a user with a Distressed ground-truth ZIP Code has a 38% chance of being geolocated to a Distressed ZIP Code. Across ground-truth tiers, the most common misidentification is to a Prosperous tier.

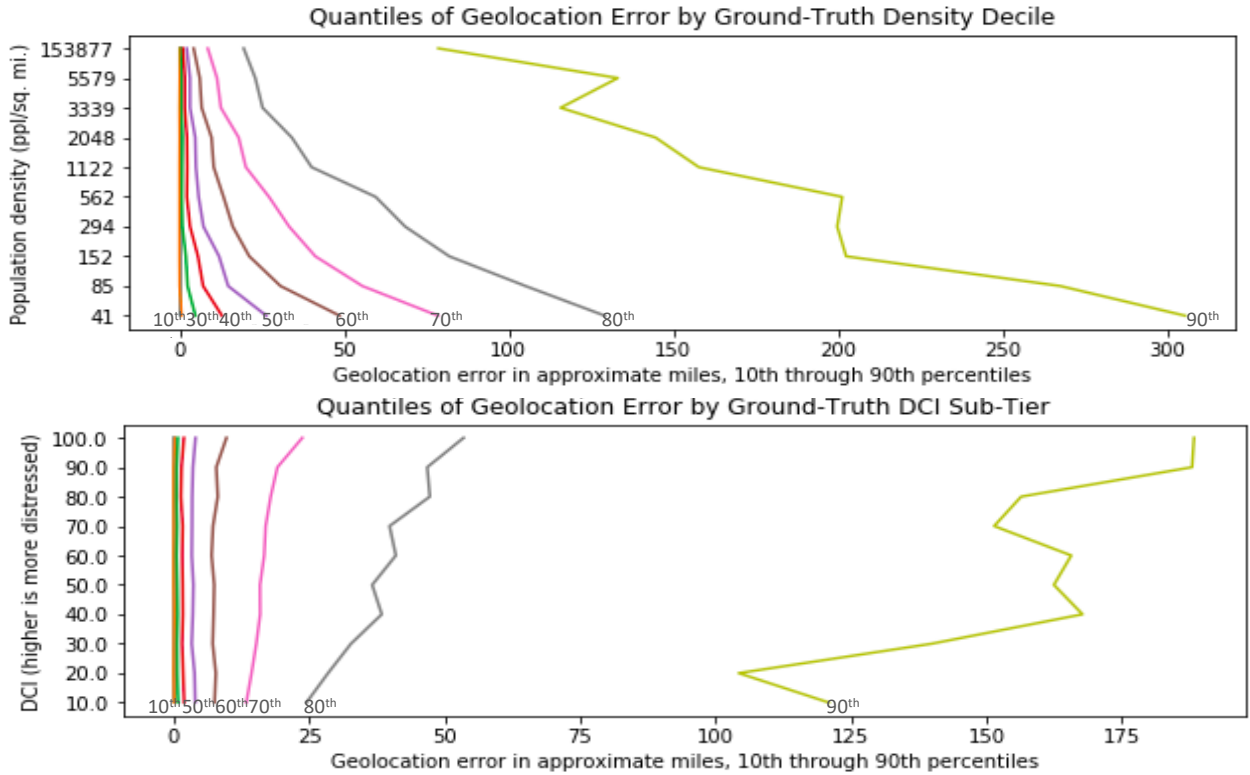

Figure 3. Percentiles of geolocation error by ground-truth density and DCI.

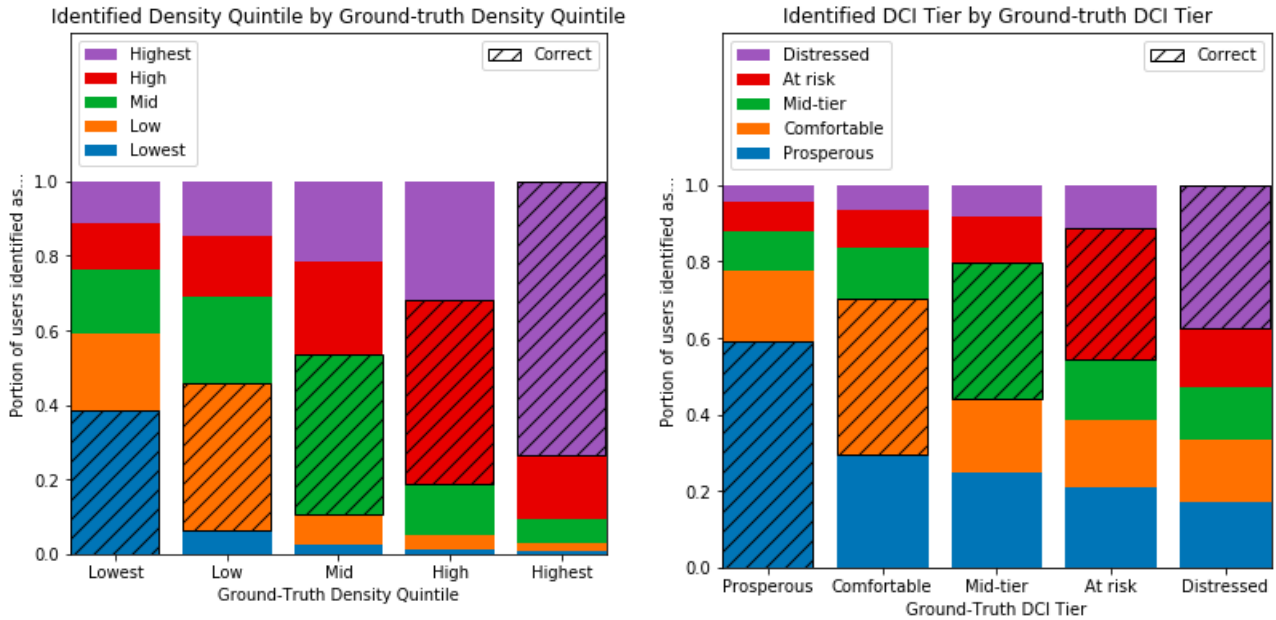

Figure 4. GeoIP ZIP Code density and DCI by ground-truth ZIP Code density and DCI

Similarly, identifications are most accurate for, and misidentifications tend to lie in the direction of, ZIP Codes of high population, low area, and high population density.

### 3.4 DCI Distributions

Regarding the overall distribution of user locations, using IP geolocation *underestimates* the regressive pattern in edX usage (as visible in the gap between red and blue curves in Figure 2b).

According to MaxMind geolocations, there are 33.2 users per 100,000 population in Prosperous areas and 16.1 in Distressed areas. According to ground-truth identifications, there are 36.2 in Prosperous areas and 14.0 in Distressed areas - a gap 30% bigger.

This is an interesting dual effect in the patterns of DCI identification. On the one hand, individual IP geolocations are biased towards more prosperous tiers: for a user chosen with equal probability from one of the five tiers, IP geolocation is more likely to identify her tier correctly if she is from a more prosperous area, and more likely to err on the side of putting her in a more prosperous than a more distressed area. On the other hand, in total the GeoIP distribution is shifted towards more distressed areas. The seeming paradox is explained by the fact that there are far more users from prosperous areas, where geolocation can err only in the too-distressed direction, than from distressed ones. In prosperous tiers, IP geolocation errs less for individual users, but errs more in total because there are more users.

## 4. CONCLUSION

### 4.1 Summary of Results

According to both IP geolocation and user-provided mailing addresses, HarvardX and MITx registrants disproportionately come from denser, more prosperous ZIP Codes.

Compared to using ZIP Codes from mailing addresses, using IP geolocation to analyze users' economic patterns produces biased results and penalizes users from more economically distressed areas. Users from less dense and more distressed areas experience larger geolocation errors, and are more likely to be geolocated to ZIP Codes with different properties than their ground-truth ones. Using IP geolocation also underestimates the regressive patterns in MOOC usage: by ground-truth data, users from prosperous areas dominate MOOCs even more than the results of IP geolocation indicate.

### 4.2 Limitations

The largest caveat to this work is that we treat user-provided mailing addresses as "ground-truth" locations. This is probably not quite right; for instance, users might report work addresses but access edX from home, move or travel, or fabricate false addresses. Nonetheless, we believe that users' mailing addresses are a substantially *more* accurate indicator of their locations than their IP geolocation, which even by MaxMind's assessment is correct at the ZIP Code level scarcely a third of the time.

Smaller limitations include occasional mistakes in address parsing and geocoding, the possibility of systematic errors or omissions in the supplementary datasets used, and misalignments in time and space between datasets.

### 4.3 Implications

Regarding economic patterns in registration, our results point in the same direction as previous research: MOOCs do not seem to be democratizing education, but rather providing more resources to people who already likely have more access to wealth, employment, and education. To reverse this pattern, MOOC developers must first understand usage in disadvantaged areas so that they can appropriately target resources.

Unfortunately, our results suggest that researchers should use great caution if relying on IP geolocation to study users' backgrounds. We find that IP geolocation not only is often inaccurate, as the networking literature suggests, but also can introduce economic bias. These biases may have affected existing MOOC research that uses fine-grained IP geolocation, such as [6], as well as other uses in academic, commercial (such as targeted advertising), and legal domains.

## 5. ACKNOWLEDGMENTS

Thanks to Mohammad Alizadeh, Hari Balakrishnan, Andrew Ho, and Joshua Goodman for feedback on this work, MIT Open Learning and Sanjay Sarma for financial support, the Harvard-MIT Data Center for computational resources, and edX and the Economic Innovation Group for data.

The findings expressed in this paper are solely those of the authors and not necessarily those of The Economic Innovation Group. The Economic Innovation Group does not guarantee the accuracy or reliability of, or necessarily agree with, the information provided herein.